\documentclass[pra,aps,showpacs,nofootinbib,preprint]{revtex4}

\begin{document}

\preprint{FERMILAB-PUB-TM-2332-A}

\title{Comments on Backreaction and Cosmic Acceleration}  

\author{Edward W. Kolb}\email{rocky@fnal.gov}
\affiliation{Particle Astrophysics Center, Fermi
       	National Accelerator Laboratory, Batavia, Illinois \ 60510-0500, USA \\
       	and Department of Astronomy and Astrophysics, Enrico Fermi Institute,
       	University of Chicago, Chicago, Illinois \ 60637-1433 USA}
\author{Sabino Matarrese}\email{sabino.matarrese@pd.infn.it}
\affiliation{Dipartimento di Fisica ``G.\ Galilei'' Universit\`{a} di Padova, 
        INFN Sezione di Padova, via Marzolo 8, Padova I-35131, Italy}
\author{Antonio Riotto}\email{antonio.riotto@pd.infn.it}
\affiliation{CERN, Theory Division, CH-1211 Geneva 23, Switzerland}

\date{\today}

\begin{abstract}
In this brief WEB note we comment on recent papers related to our paper 
{\it On Acceleration Without Dark Energy} \cite{us}.
\end{abstract} 


\maketitle

In Ref.\ \cite{us} we elaborated on the proposal that the observed acceleration
of the Universe might be the result of the backreaction of sub-horizon
cosmological perturbations, rather than the effect of a negative-pressure
dark-energy fluid or a modification of general relativity.  Through studying
the effective Friedmann equations describing an inhomogeneous Universe after
smoothing (see the work of Buchert \cite{buchert}), we suggested that
acceleration in our Hubble volume might be possible even if local fluid
elements do not individually undergo accelerated expansion.

To describe the time evolution of a region of the Universe as large as our
local Hubble volume one has to construct the effective dynamics from which
observable average properties can be inferred. This is intimately connected
with the general problem of how the (possibly nonlinear) dynamics of
cosmological perturbations on small scales affects the large-scale
``background'' geometry, and with the process of averaging over a given domain
${\cal D}$ of volume $V_{\cal D}$.

The scale factor averaged over a domain ${\cal D}$ is defined by $a_{\cal
D}\equiv \left(V_{\cal D}\right)^{1/3}$. The very simple fact that the 
averaging of the time derivative of a locally defined quantity differs 
from the time derivative of
the averaged quantity implies that acceleration is possible in principle for
the dynamics described by the average scale factor $a_{\cal D}$.

Indeed, from the effective equations of motion, it is easy to show that
acceleration may be achieved if $Q_{\cal D}>4\pi G\langle\rho\rangle_{\cal D}$,
where $Q_{\cal D}$ is the kinematical backreaction term encoding the effect of
nonlinearities \cite{us,buchert} and $\rho$ the local energy density.

Nambu and Tanimoto \cite{nambu}, propose an explicit example of an
inhomogeneous Universe that leads to accelerated expansion after taking spatial
averaging.  The model contains both a region with positive spatial curvature
and a region with negative spatial curvature. It was found that after the
region with positive spatial curvature begins to re-collapse, the deceleration
parameter of the spatially averaged Universe becomes negative and the averaged
Universe starts an accelerated expansion phase. While this is not represented
as a realistic cosmological model, it illustrates several important concepts.

Generalizing the model of Ref.\ \cite{nambu}, to understand how backreactions
can lead to acceleration, one can think of the domain ${\cal D}$ as a
collection of smaller regions which are individually homogeneous and isotropic
and expanding with a scale factor $a_i$ and rate $H_i$. The corresponding
average acceleration in this case is given by
\begin{eqnarray}
\frac{\ddot{a}_{\cal D}}{a_{\cal D}} & = & 
\left\langle \frac{\ddot{a}(x)}{a}\right\rangle 
+ 2 \left[ \left\langle \left(\frac{\dot{a}(x)}{a(x)}\right)^2 \right\rangle 
- \left(\left\langle \frac{\dot{a}(x)}{a(x)} \right\rangle\right)^2 \right] 
\nonumber \\ 
& = & \sum_i
\frac{\ddot{a}_i}{a_i}\frac{V_i}{a^3_{\cal D}}  + 2 \sum_i \left(
\frac{\dot{a}_i}{a_i}\right)^2\frac{V_i}{a^3_{\cal D}} - 2 \left( \sum_i \left(
\frac{\dot{a}_i}{a_i}\right) \frac{V_i}{a^3_{\cal D}} \right)^2 
\end{eqnarray} 
where $V_i$ is the volume of region $i$.  A necessary (but not sufficient)
condition for getting acceleration is that some of the regions evolve with a
different Hubble rate than others. In Ref.\ \cite{wald} this strategy was
criticized on the basis that one can envisage situations in which, upon
averaging, the Universe accelerates, despite the fact that the observers of
each individual separate region experience deceleration $(\ddot{a}_i<0)$.

Indeed, Ishibashi and Wald \cite{wald} consider the case in which there are two
of such separate regions for which $a_1=a_2$ and $\dot{a}_1=-\dot{a}_2$ (how
these conditions may be preserved by the dynamics, {\it e.g.}, in the model of
Ref.\ \cite{nambu} is not clear to us). They claim that this ``graphically
illustrates'' that the requirement $\ddot{a}_{\cal D}>0$ is ``far from
sufficient to account for the physically observed effects of acceleration in
our universe.'' But our basic point is {\it exactly} that acceleration of the
mean scale factor can occur even though individual elements decelerate.  The
fact that $a_{\cal D}$ is related to observables is strongly suggested by the
work of Tomita \cite{tomita}, who calculated observables such as the luminosity
distance and the angular-diameter distance in an inhomogeneous models and finds
apparent acceleration.

The authors of Ref.\ \cite{wald} also argue that the averaging procedure is
affected by ambiguity both in regard to the choice of time slicing and the
choice of the domain ${\cal D}$. However, the statement about the time slicing
is equivalent to say that whether acceleration is experienced or not depends
upon the observer. This is true even for an unperturbed
Friedmann-Robertson-Walker (FRW) spacetime in which only those observers
comoving with the perfect fluid source would say that 
the Universe is homogeneous
and isotropic. We do not see any ambiguity in choosing the observer comoving
with the matter flow.  The dependence of the average parameters on the choice
of the domain ${\cal D}$ is an unavoidable consequence of the standard
procedure of fitting an FRW model to a real perturbed Universe. That said, the
real issue is the appropriate scale over which inhomogeneities need to be
smoothed out, given the specific dataset one wishes to fit. For instance,
Tomita has investigated the possibility that we live in a locally underdense
region of size about $(200-300)$ Mpc and studied the magnitude-redshift
relation in a cosmological model with such a local void \cite{tomita}.  The
accelerating behavior of high redshift supernovae can be explained in this
model, because the local void plays a role similar to the positive cosmological
constant. Acceleration is experienced by the observer living inside the void as
her/his region expands faster than the outer region despite 
the fact that both are
decelerating. In the average language, this situation precisely results in the
simultaneous presence of largely under-dense and over-dense regions giving rise
to a large kinematical backreaction \cite{us}.  Notice also that the dependence
over the volume $V_{\cal D}$ disappears and one can safely replace the spatial
average over ${\cal D}$ with the ensemble average as soon as the volume is
large enough for the Ergodic Theorem to hold.

Another criticism raised in Ref.\ \cite{wald} is that if the Universe is
accurately described by a Newtonian perturbed FRW metric, then  no backreaction
may give rise to acceleration.  We fully agree on the fact that Newtonian
approximation yields an  accurate description of our Universe on all relevant 
scales (as long as the considered wavelength is much larger than the 
Schwarzschild radius of collapsing  bodies) as we explicitly 
emphasized in Section IIIC of Ref.\ \cite{us}. 
The real issue is which quantity should be computed for a proper
evaluation of the impact of the backreaction. Acceleration requires the
kinematical backreaction term to be of the same order of the average curvature
of comoving hypersurfaces. The latter is well known to vanish at the Newtonian
level, but it nevertheless enters the dynamics of the Universe.  One needs
therefore a genuine relativistic description. In the Newtonian case, it is
immediate to verify that $Q_{\cal D}$ is {\it exactly} ({\it i.e.,} at any
order in perturbation theory) given by the volume integral of a
total-derivative term in Eulerian coordinates, so that by the Gauss theorem it
can be transformed into a pure boundary term.  It is precisely for this reason
that any consideration of the backreaction based on the Newtonian approximation
is not helpful: it will invariably lead to a tiny effect, and to the absence of
any acceleration.  What is important for us here is that $Q_{\cal D} $ clearly
displays sizeable non-Newtonian terms even in the weak-field gauge. 
The key point is that the backreaction has to be calculated adopting the
proper time of observers comoving with the matter flow. It is precisely
for this reason that sizeable post-Newtonian backreaction terms, of the
type $\sim H^2 \langle \delta^2 (v/c)^2\rangle$ (where $\langle \cdot \rangle$ 
stands for the ensemble average), appear in the effective
Friedmann equations describing an inhomogeneous Universe after smoothing.
Notice that this occurs {\it in spite of the fact} that in the Poisson
gauge the metric itself is very well approximated by the weak-field form,
{\it i.e.} Eq.\ (77) of Ref.\ \cite{us} with $\phi_P=\psi_P \equiv
\Phi_N/c^2 \ll 1$. (Here the subscript $N$ stands for Newtonian.  Our form
is, of course, identical to Eq.\ (1) of Ref.\ \cite{wald}, where $\Psi
\equiv \Phi_N/c^2$).
We take the opportunity to stress once again that in the weak-field 
approach the number of gradients is finite and the complexity of the 
problem resides in the non-perturbative
evaluation of the evolved potentials in terms of the initial seeds. The
situation is reversed when approaching the problem in the synchronous and 
comoving gauge, where the expressions for backreaction terms have to 
be expanded through an infinite series of gradients of the initial seed itself.

A  possible objection to the use of the synchronous and  comoving gauge in
addressing the backreaction problem is  the occurrence of shell-crossing
singularities (caustics) in the evolution of collisionless fluids, which 
might prevent  the analysis to be carried over into the fully non-linear
regime.  We would like to point out that the instability we find in Ref.\
\cite{us} in the gradient expansion is unrelated to  shell-crossing
singularities.  This can be immediately appreciated by noting that: i)
shell-crossing instabilities imply the emergence of {\it divergent} gradients
terms, while our instability shows up through an infinite number of {\it
finite} gradient terms; ii) shell crossing is well known to lead to an infinite
Newtonian term, while our effect involves a tiny Newtonian term.  It should
also be stressed that the occurrence of caustics does not represent a serious
limitation of our approach; indeed, the very fact that caustics only carry a
small amount of mass implies that they can be easily smeared over a finite
region out in such a way that their presence does not affect the mean expansion
rate of the Universe.

\acknowledgments
E.W.K.\ is supported in part by NASA grant NAG5-10842
and by the Department of Energy. S.M. ackowledges partial financial support 
by INAF. A.R. is on leave of absence from INFN, Sezione di Padova, Italy. 
We would like to acknowledge Jim Bardeen and Bob Wald for useful discussions. 


\end{document}